# Secular Aberration Drift and IAU Definition of ICRS

Oleg Titov

*Geoscience Australia, PO Box 378, Canberra, ACT 2601, Australia*

ABSTRACT

The gravitational attraction of the Galactic centre leads to the centrifugal acceleration of the Solar system barycentre. It results in secular aberration drift which displaces the position of the distant radio sources. The effect should be accounted for in high-precision astrometric reductions as well as by the corresponding update of the ICRS definition.

## 1 INTRODUCTION

Conventional reductions of high-precision Very Long Baseline Interferometry (VLBI) data are undertaken in the reference system with an origin at the barycentre of Solar system. This barycentre reference system was adopted by the International Astronomical Union as the International Celestial Reference System (ICRS). By definition, the axes of the ICRS are the adopted positions of a specific set of extragalactic objects (i.e. very distant quasars), *which are assumed to have no measurable proper motions* (McCarthy & Petit 2004). The ICRS axes are consistent, to better than 0.1 arcsec, with the equator and equinox of J2000.0 defined by the dynamics of the Earth. To realize the ICRS, a set of fiducial sources with their coordinates needs to be established in the form of a catalogue. From 1998 till 2009 the International Celestial Reference Frame (ICRF) was a catalogue of 608 radio source positions observed with VLBI, of those 212 were so-called defining sources used to establish the orientation of the ICRS axes (Ma et al. 1998). In 2009 this first ICRF realization (ICRF1) was replaced by the ICRF2, which included positions of 3414 radio sources (Fey, Gordon & Jacobs 2009). The set of the ICRF2 defining sources contains 295 objects, and only 97 of them were defining sources in the ICRF1. The formal accuracy of the new ICRF2 realization has been improved, therefore the assumption mentioned above should be properly checked with the modern VLBI results. Particularly, the Galactocentric acceleration of the Solar system barycentre is not considered as a part of the ICRS definition, and may produce a measurable systematic effect (secular aberration drift) in the apparent displacement of the reference radio sources (Bastian 1995; Eubanks et al. 1995; Gwinn et al. 1997; Sovers, Fanselow and Jacobs 1998, Kovalevsky 2003; MacMillan 2005; Kopeikin & Makarov 2006). This effect has not been found so far, possibly, due to inadequate analysis methodology used for analysis of VLBI data. In this Letter, we propose two alternative approaches to detect the secular aberration drift.

## 2 FORMAL ACCURACY OF THE ICRF

The formal accuracy $\sigma_0$ of the most observed radio source positions in the ICRF2 is about 7 μas in both components (Goddard Space Flight Centre (GSFC) solution gsfc2008a). However, it was found that the realistic ('inflated') accuracy $\sigma_{inf l}$ is essentially larger. The inflated accuracy is calculated by the following formula:

$$\sigma_{inf l}^2 = (1.5\sigma_0)^2 + (0.04 mas)^2 \quad (1)$$

Thus, for the most observed radio sources $\sigma_{inf l}$ equal to 41.4 μas (Fey, Gordon & Jacobs 2009) and the ratio $\frac{\sigma_{inf l}}{\sigma_0} \approx 6$. For comparison, the analogous formula for the ICRF1 realization is given by (Ma et al. 1998)

$$\sigma_{inf l}^2 = (1.5\sigma_0)^2 + (0.25 mas)^2$$

For ICRF1 the minimum inflated error was $\sigma_{inf l} \approx$ 0.26 mas, and with the $\sigma_0 \approx$ 0.06 mas the ratio $\frac{\sigma_{inf l}}{\sigma_0} \approx 4$ is similar to the ratio for ICRF2 (1).

We believe that this disagreement between formal and inflated errors could be caused by hidden systematic effects. Herewith, one of possible systematic effects, secular aberration drift, is discussed.

## 3 THE GALACTOCENTRIC ACCELERATION AND THE SECULAR ABERRATION DRIFT

In accordance with the ICRS definition, the barycentre reference system is assumed to be quasi-inertial, i.e. the net rotation of the reference quasars is negligible. This assumption is reasonable because all the distant quasars are unlikely to have common motion in the Universe, at least on the current level of observational accuracy. However, the definition of the quasi-inertial reference system permits accelerated motion of the origin with respect to the reference points (Hofmann-Wellenhof et al. 2003). The Galactocentric rotation of the Solar system barycentre around along with other stars results in the systematic apparent proper motion (secular aberration drift) towards to the Galactic centre (the apex point) with the equatorial coordinates ($\alpha_G, \delta_G$). The proper motion of an object with equatorial coordinates ($\alpha, \delta$) is given by (Kopeikin & Makarov 2006)

$$\mu_\alpha \cos\delta = \frac{1}{c}(-a_1 \sin\alpha + a_2 \cos\alpha)$$
$$\mu_\delta = \frac{1}{c}(-a_1 \sin\delta\cos\alpha - a_2 \sin\delta\sin\alpha + a_3 \cos\delta) \quad (2)$$

where $(a_1, a_2, a_3)$ - components of the vector of the Galactocentric acceleration $A$ converted to the units of arcsec yr$^{-1}$ and calculated as

$$a_1 = \mathbf{A}\cos\alpha_G \cos\delta_G, a_2 = \mathbf{A}\sin\alpha_G \cos\delta_G, a_3 = \mathbf{A}\sin\delta_G$$

The theoretical Galactocentric acceleration [magnitude of $2\times10^{-13}$ km s$^{-2}$ (Walter & Sovers (2000)] corresponds to the maximum systematic motion of 4 μas y$^{-1}$ on the angular distance 90º from the apex point. As it was stated by Bastian (1995): 'Thus, all quasars will exhibit a distance-independent streaming motion towards the galactic centre'. Over a 20-yr time span of the highest precision VLBI data, the total displacement of the radio sources reaches 80 µas and, thus, substantially exceeds the formal error of the most observed radio sources. Therefore, this dipole effect of secular aberration drift, not included in the reduction models recommended by the IERS Conventions 2003 (McCarty & Petit 2004), might cause a large shift between the actual and catalogue coordinates. Thus, orthogonality of the ICRS axes fixed by the position of reference radio sources might be lost.

## 4 DISCUSSION OF THE PROCEDURES FOR VLBI DATA ANALYSIS

The procedure of radio source position estimation is complicated due to effect of source structure. The motion of super-relativistic jets causes apparent motion of radio sources with rates of up to 500-1000 µas yr$^{-1}$ (Charlot 1990; Johnston et al. 1995; MacMillan & Ma 2007; Titov 2007) on short time scales (from several months to several years), though, they are typically in the range of 50-100 µas yr$^{-1}$. To minimize the effect of intrinsic instability, the most unstable radio sources are rejected from the list of candidates. Then positions of the selected radio sources are considered as relatively stable over the 10-20-yr time interval and estimated as 'global' parameters with no-net-rotation constraints (NNR), i.e. only one estimate of coordinates over the 10-20 yr period is produced (Ma et al. 1998; Feissel-Vernier 2003). Positions of the 'unstable' radio sources are either approximated by a piecewise linear function (MacMillan & Ma 2007) or estimated for each 24-h VLBI session separately (Titov 2007). For the second scenario, mean positions and proper motion of the unstable radio sources can be calculated from the time-series of daily coordinates. However, these mean positions are embedded by the fixed positions of the reference radio sources. Once the systematic effects are not permitted among the reference radio sources, they are not also permitted among the unstable radio sources.

Up until the mid-1990s no systematic effects were found (Walter & Sovers 2000), and the assumption about smallness of systematic effects in proper motion remained valid. However, due to the dramatic improvement of the VLBI data precision over the last 15 yr and the consequent improvement in the precision of astrometric positions, the negligibility of the systematic does not hold anymore. Thus, the conventional approach effectively suppresses any existing systematic, such as the secular aberration drift, along with the fictitious systematic due to intrinsic instability of the individual source positions.

One could propose two alternative methods for estimation of the radio source positions

1. No radio sources positions should be treated as 'global' parameters. This means that all positions of radio sources are estimated for each 24-h session, and any existing systematic is not suppressed by this approach. Time-series of the radio source daily positions are produced. By fitting the time-series by a linear function we could estimate the linear rate (proper motion) of all sources having a sufficient number of observations. Eventually, the components of the secular aberration drift are estimated by several hundred individual proper motions. This approach has been realized by Gwinn et al (1997).

2. Positions of the reference radio sources are treated as 'global' parameters following the conventional approach. However, three components of the secular aberration drift ($a_1, a_2, a_3$) from (2) are to be estimated as 'global' parameters as well. Thus, the least-squares parametrical model is not incomplete because the three parameters absorb the secular aberration drift systematic effect, which is estimated without intermediate calculations of the proper motion. This approach has been realized by MacMillan (2005) and Titov (2009).

Both approaches are practically equivalent. In the first approach the systematic motion (2) of the reference radio sources is introduced explicitly. For the second option, the Galactocentric acceleration of the Solar system barycentre should be incorporated into the geocentric vacuum delay model as recommended by IERS (McCarthy & Petit 2004) instead of introducing the systematic proper motion effect. This practical approach leads to the theoretical discussion on the ICRS definition. It may be more preferable to modify the conventional metric tensor in such a way as to keep the Solar system barycentre as having no acceleration with respect to the Galactic centre. However, this theoretical work is beyond the scope of this Letter.

Due to uneven distribution of the reference radio sources around the sky, the estimates of the vector spherical harmonics of the first and second degree are highly correlated (Titov & Malkin 2009). As a result, the parameters of the secular aberration drift may be coupled with the other signals, especially, the component $a_3$ because this is estimated only from one component of the proper motion $\mu_\delta$ (2).

More observational programmes should be organized now in the Southern hemisphere to increase the amount of measured proper motion and, subsequently, to reduce the correlation between estimated parameters. Three 12-m radio telescopes in Australia (the AuScope project) and a 12-meter radio telescope in New Zealand (hosted by Auckland University of Technology) will start operation in second half of 2010 (Titov et al 2008; Tregoning et al. 2008). These new facilities will be used for high-precision astrometry of the reference radio sources.

5 CONCLUSION

The definition of the ICRS considers the barycentre as an isolated point without external mass beyond the Solar system boundaries. However, the barycentre is in a state of free-fall in the gravitational field of the Galaxy. This satisfies to the definition of a quasi-inertial reference system (Hofmann-Wellenhof et al, 2003), but as such the ICRS cannot be considered as an inertial reference system. This concept of quasi-inertial reference system was focused on removing a rotation of the ICRS axes. However, the dipole systematic effect would result in violation of orthogonality of the ICRS axes, even if their rotation is zero.

The effect of secular aberration drift may cause a dipole systematic proper motion described by equation (2) with a magnitude of 4 μas yr$^{-1}$, leading to a total displacement of 80 μas for a 20-yr period of data collection, which is much larger than the formal and inflated positional errors in the ICRF2. Neglecting these factors may result in a bias of other parameters estimated from VLBI data and to lead to a discrepancy between the formal and realistic accuracy of the radio source coordinates (1). Therefore, it is necessary either to estimate the additional parameters during adjustment of VLBI data or revise the related reductional models.

ACKNOWLEDGEMENTS

I thank Sergei Kopeikin, Sergei Klioner, John Dawson and the anonymous reviewer for useful suggestions.

REFERENCES

Bastian U., 1995, in Perryman A. C., van Leeuwen, F., ed. Proc. of the RGO-ESA Workshop Future Possibilities for Astrometry in Space, ESA SP-379. EAS, Noordwijk, p. 99.

Charlot P., 1990, AJ, 160, 1309

Gwinn C. R., Eubanks T. M., Pyne T., et al., 1997, ApJ, 485, 87

Eubanks T. M. et al., 1995, in Hog E., Seidelmann K., eds., Proc IAU Symp. 166, Astronomical and Astrophysical Objectives of Sub-milliarcsecond Optical Astrometry. Kluwer, Dordrecht, p. 283

Feissel-Vernier M., 2003, A&A, 403, 105

Fey A., Gordon D., Jacobs C., eds, 2009, IERS Technical Note 35, The Second Realization of the International Celestial Reference Frame by Very Long Baseline Interferometry

Gwinn C. R., Eubanks T. M., Pyne T., Birkinshaw M., Matsakis D. N., 1997, ApJ, 485, 87

Hofmann-Wellenhof B., Legat K, Wieser M., Lichtenegger H., 2003, Navigation, Springer, NewYork

Johnston K. J. et al., 1995, AJ, 110, 880

Kopeikin S., Makarov V., 2006, AJ, 131, 1471

Kovalevsky J., 2003, A&A, 404, 743

Ma C. et al., 1998, AJ, 116, 516

MacMillan D., 2005, in Romney J. D., Reid M., eds, Future Directions in High Resolution Astronomy: The 10$^{th}$ Anniversary of the VLBA, Astron. Soc. Pac., San Francisco, p. 477

MacMillan D., Ma C., 2007, J. Geodesy, 81, 443

McCarthy D.D., Petit G., 2004, IERS Conventions 2003, Verlag des Bundesamtes fur Kartographie und Geodasie, Frankfurt am Main

Sovers O. J., Fanselow J. L., Jacobs C. S., 1998, Rev. Modern Phys., 70, 1393

Titov O., 2007, Astron Lett., 33, 481

Titov O., 2009, in Bourda G., Charlot P., Collioud A., eds., Proc 19th European VLBI for Geodesy and Astrometry Working Meeting, Univ. de Bordeaux, Bordeaux, p. 14

Titov O., Malkin Z., 2009, A&A, 506, 1477

Titov O., Gulyaev S., Lovell J., Dickey J., 2008, in Finkelstein A., Behrend D., eds., Proc 5th IVS General Meeting, Nauka, Saint Petersburg, p. 114

Tregoning et al., (The University Component of the AuScope Geospatial Team), 2008, J. Spatial Sci., 53, 65

Walter H. G., Sovers O. J., 2000, Astrometry of Fundamental Catalogues: the Evolution from

Optical to Radio Reference Frames, Springer-Verlag, Berlin